\documentclass[%
 reprint,
 amsmath,amssymb,
 aps,
]{revtex4-1}

\usepackage{graphicx}
\usepackage{dcolumn}
\usepackage{bm}

\begin{document}

\preprint{APS/123-QED}

\title{An explicit formula for the polynomial entanglement measures of degree 2 of even-N qubits mixed
states}

\author{M.A. Jafarizadeh}
 \altaffiliation{jafarizadeh@tabrizu.ac.ir}
\author{M. Yahyavi}%
 \email{m.yahyavi@tabrizu.ac.ir}
\affiliation{%
 Department of Theoretical Physics and Astrophysics, Tabriz University, Tabriz 51664, Iran.\\
}%

\author{A. Heshmati}
 \email{heshmati@iaushab.ac.ir}
\affiliation{
Department of Physics, Shabestar Branch, Islamic Azad University, Shabestar, Iran.\\
}%

\author{N. Karimi}
\email{n.karimi@cfu.ac.ir}
\address{Department of Science, Farhangian University, Tehran 19989-63341, Iran.}
\author{A. Mohamadzadeh}
\email{a.mohamadzadeh@tabrizu.ac.ir}
\address{Department of Theoretical Physics and Astrophysics, Tabriz University, Tabriz 51664, Iran.}
\author{F. Eghbalifam}
\email{F.Egbali@tabrizu.ac.ir}
\address{Department of Theoretical Physics and Astrophysics, Tabriz University, Tabriz 51664, Iran.}
\author{S. Nami}
\email{S.Nami@tabrizu.ac.ir}
\address{Department of Theoretical Physics and Astrophysics, Tabriz University, Tabriz 51664, Iran.}

\date{\today}

\begin{abstract}
Characterization of the multipartite mixed state entanglement is still a challenging problem. Since due to the fact that the entanglement for the mixed states, in general, is defined by a convex-roof extension.
That is the entanglement measure of a mixed state $\rho$ of a quantum system can be defined as the minimum average entanglement of an ensemble of pure states. In this Letter, we give an explicit formula for the  polynomial entanglement measures of degree 2 of even-N qubits mixed states that is similar to Wootters formula in \cite{Wootters}. Then we discuss our
findings in the framework of X density matrices and show that our formula for this type of density
matrices is in the full agreement with the genuine multipartite (GM) entanglement of these states.

{\it {\bf PACS numbers:}} 03.67.Mn, 03.65.Ud.

\end{abstract}

\pacs{Valid PACS appear here}
\maketitle


\section{\label{sec:level1}introduction}

A fundamental element of quantum information science is quantum entanglement. Quantum entanglement is a physical phenomenon that occurs when pairs or groups of particles interact in ways such that the quantum state of each particle cannot be described independently of the others. Such phenomena were the subject of a 1935 paper by Einstein, Podolsky, and Rosen\cite{Einstein} and several papers by Schrödinger shortly thereafter like \cite{Schrödinger} describing what came to be known as the EPR paradox. Quantum entangled states are crucial resource and play key roles in quantum information processing such as quantum teleportation \cite{teleportation}, quantum cryptography \cite{cryptography} and quantum computation \cite{computation}. An entangled system is defined to be one whose quantum
state cannot be factored as a product of states of its local
constituents. If a pure state $|\psi\rangle\in H_{1}\otimes H_{2}\otimes...\otimes H_{n}$- $H_{i}$ is the Hilbert space of the i-th subsystem- can be written in the form $|\psi\rangle=|\psi_{1}\rangle\otimes|\psi_{2}\rangle\otimes...\otimes|\psi_{n}\rangle$   where $|\psi_{i}\rangle$  is a pure state of the i-th subsystem, it is said to be separable, otherwise it is called entangled. When a system is in an entangled pure state, it is not possible to assign states to its subsystems.This will be true, in the appropriate sense, for the mixed state case as well. A mixed state of the composite system is described by a density matrix $\rho$  acting on $H_{1}\otimes H_{2}\otimes...\otimes H_{n}$. $\rho$ is separable if there exist $p_{k}\geq 0$, $\{\rho _{1}^{k}\}$, $\{\rho _{2}^{k}\}$,..., $\{\rho _{n}^{k}\}$ which are mixed states of the respective subsystems such that $\rho =\sum _{k}p_{k}\rho _{1}^{k}\otimes \rho _{2}^{k}\otimes...\otimes\rho_{n}^{k}$ where $\;\sum _{k}p_{k}=1.$\\
So far, different measures of entanglement for mixed quantum states is introduced like the entanglement of formation \cite{Benett}, the entanglement cost \cite{Hayden} and the distillable entanglement \cite{Benett}. All of these measures are common in the some properties: They arrive to zero for each separable state, are invariant under local unitary transformations, and are never increasing on average by
local operations and classical communication (LOCC) \cite{Vedral1, Vedral2, Eltschka}. The latter property meaning that the entanglement measure is a so-called entanglement
monotone. One of the ways to determining the entanglement of a pure quantum states is a polynomial function in the coefficients of states which are
invariant under stochastic LOCC (SLOCC) and play a critical role in the investigation
of entanglement measures. The polynomial function $P$ of degree $l$ of a system of $m$ qudits is defined as:\\
$$P(\kappa L|\psi\rangle)=\kappa^{l}P(|\psi\rangle),$$
for a constant $\kappa > 0$ and an invertible linear operator $L\in
SL(l, C)^{\times m}$ representing the SLOCC transformation. The
absolute value of any such polynomial with $l\leq 4$ defines in
fact an entanglement monotone \cite{Bastin}. For
two and three qubits, the concurrence and the three-tangle are
polynomial invariants of degrees 2 and 4, respectively \cite{Wootters, Coffman}. Many
efforts have been done over the last decade on the
study of polynomial invariants for four or more qubits
\cite{Wong, Osterloh, Li, Viehmann}. \\
A polynomial invariant E is extended to mixed states
by means of the convex roof that  is the largest convex
function on the set of mixed states which corresponds to E on
pure states, given for a mixed state
matrix $\rho$ by:
$$E(\rho)=min_{\{p_{i},|\psi_{i}\rangle\}}\sum_{i}p_{i}E(\psi_{i}),$$
where $\{p_{i},|\psi_{i}\rangle\}$ is the ensemble of a pure state for the given density matrix $\rho$ such that $\sum_{i}p_{i}=1$. The ensemble that minimizes $E(\rho)$ is called optimal. Any convex hull  of
all pure states that is vanished E, called zero-polytope and in the outside of this convex hull, it will never vanish. \\
 In \cite{Wootters}, Wootters find an explicit formula for the entanglement of formation of a pair of qubits as a function of their density matrix. In this paper, we first explain the concept of polynomial invariant of degree 2 for even-N qubits of pure states. The absolute value of this invariant is entanglement monotones. Then following Wootters \cite{Wootters}, we find the exact formula of this polynomial invariant of degree 2 for any even-qubits density matrix.

\section{ Polynomial invariant of degree 2}
A function that quantify the entanglement of quantum states must be non-increasing (on average) under stochastic
local operations and classical communication (SLOCC) where is so-called an entanglement monotone \cite{Vidal}.

Here we investigate the polynomial invariant of degree 2 as an entanglement measure for any even-N qubit quantum states.
Generic pure N-qubit states are of the form
\begin{eqnarray}\label{qubit}
    |\psi_{A_{1}A_{2}...A_{N}}\rangle=\sum_{i_{1},i_{2},...,i_{N}=0}^{1}\psi_{i_{1}i_{2}...i_{N}}|i_{1}i_{2}...i_{N}\rangle,
\end{eqnarray}\\
where , due to normalization, $\sum_{i_{1},i_{2},...,i_{N}=0}^{1}|\psi_{i_{1}i_{2}...i_{N}}|^{2}=1$. For any even-N qubits pure quantum state, the degree-2 invariant is defined as:
\begin{eqnarray}\label{multi-qubit invariant}
C(\left| {{\psi _{{A_1}...{A_N}}}} \right\rangle ) = {\varepsilon _{{i_1}{j_1}}}{\varepsilon _{{i_2}{j_2}}}...{\varepsilon _{{i_N}{j_N}}}
{\psi _{{i_1}{i_2}...{i_N}}}{\psi _{{j_1}{j_2}...{j_N}}},
\end{eqnarray}
where summation is over the repeated indices that values are 0 and 1, and $\varepsilon$ is the SL(2,C)-invariant alternating tensor
$$\varepsilon:=\left(
                 \begin{array}{cc}
                   0 & 1 \\
                   -1 & 0 \\
                 \end{array}
               \right).
$$ \\
Note that for any odd-qubits pure quantum state, this invariant is zero.\\

As the same way,
given a density matrix $\rho$ of the quantum systems with N-qubits that N is even, consider all ensembles of pure states $|\psi_{i_{1},i_{2},...,i_{N}}\rangle$ with probabilities $p_{i_{1},i_{2},...,i_{N}}$, such that
\begin{eqnarray}\label{mixed state}
   \rho=\sum_{i_{1},i_{2},...,i_{N}=0}^{1}p_{i_{1},i_{2},...,i_{N}}|\psi_{i_{1},i_{2},...,i_{N}}\rangle\langle\psi_{i_{1},i_{2},...,i_{N}}|,
\end{eqnarray}
where for any pure state, the polynomial invariant of degree 2, $C(\psi_{i_{1},i_{2},...,i_{N}})$, is defined as the Eq(\ref{multi-qubit invariant}). The degree 2 invariant of this mixed state $\rho$ is obtained by the convex roof concept that the degree 2 invariant defined first on the set of pure states and then extended to the set of all mixed states by minimizing its
average value over all possible convex decompositions of the given state $\rho$ into pure states
\cite{Benett}:
\begin{widetext}
\begin{equation}\label{mixed concurrence}
C(\rho)=min_{\{p_{i_{1},i_{2},...,i_{N}},|\psi_{i_{1},i_{2},...,i_{N}}\rangle\}}\sum_{i_{1},i_{2},...,i_{N}=0}^{1}p_{i_{1},i_{2},...,i_{N}}C(\psi_{i_{1},i_{2},...,i_{N}}).
\end{equation}
\end{widetext}
The decomposition(s) $\{p_{i_{1},i_{2},...,i_{N}},|\psi_{i_{1},i_{2},...,i_{N}}\rangle\}$ realising the minimum value of Eq(\ref{mixed concurrence}), is (are) called optimal. Wootters in \cite{Wootters} showed how to find the optimal decompositions for the most simple
bipartite cases where enables us to compute the concurrence,
analytically for the arbitrary two-qubits mixed states. In \cite{Regula},  the problem of determining the amount of entanglement of rank-2 state with any polynomial entanglement measure is seen as a geometric problem on the corresponding
Bloch sphere. In \cite{Uhlmann}, Osterloh and co-workers provided a non-trivial lower bound for the
convex roof  by means of the two concepts: zero-polytope and convex characteristic curve.  In \cite{Uhlman2, Cao, Eltschka2, He, Jang} found the explicit expressions for
the three-tangle and optimal decompositions for three-qubit mixed states of rank-n(n=2,3,...,8) examples. In the next section of this letter, we offer the exact formula of the polynomial invariant of degree 2 for any even qubit mixed state.
\section{Exact formula for the even-qubit density matrixs}
In \cite{Wootters}, Wootters found the exact formula for an arbitrary state of two qubits by means of the measure of entanglement is called the entanglement of formation. In this letter, we obtain a similar formula for any even-qubit quantum system by means of degree-2 polynomial invariant.\\
Let us first consider a pure state $|\psi\rangle$ of even-N qubit state. The degree-2 invariant $C(\psi)$ of this state is defined to be $C(\psi)=|\langle\psi|\widetilde{\psi}\rangle|$, where the tilde represents the "spin-flip" operation $|\widetilde{\psi}\rangle=(\underbrace{\sigma_{y}\otimes\sigma_{y}\otimes...\otimes\sigma_{y}}_{N-times})|\psi^{*}\rangle$. Here $|\psi^{*}\rangle$ is the complex conjugate of $|\psi\rangle$ in the standard basis $\{|0...0\rangle,...,|1...1\rangle\}$, and $\sigma_{y}$ is the Pauli operator $\left(
   \begin{array}{cc}
     0 & -i \\
     i & 0 \\
   \end{array}
 \right)$. Similarly, for a general state $\rho$ of even-N qubits, the spin-flipped state is
\begin{eqnarray}\label{spin-flipped}
   \widetilde{\rho}=(\underbrace{\sigma_{y}\otimes\sigma_{y}\otimes...\otimes\sigma_{y}}_{N- times})\rho^{*}(\underbrace{\sigma_{y}\otimes\sigma_{y}\otimes...\otimes\sigma_{y}}_{N- times}),
\end{eqnarray}
where again the complex conjugate is taken in the standard basis.\\
 We express the main result of this letter that is the exact formula for the polynomial invariant of any even-N qubits mixed state as the following theorem:\\\\\\\\\\\\\\\\\\\\\\\\\\\\\\
{\bf Theorem:
The polynomial entanglement measure of degree-2  for any even-N qubits mixed state $\rho$ in the binary form is given by:
\begin{eqnarray}\label{HHHHHH}
  {\bf C(\rho)=max\{0,\lambda_{0,0,...,0}-\sum_{i_{1},i_{2},...,i_{N}=0}^{1}\lambda_{i_{1},i_{2},...,i_{N}}\},}
\end{eqnarray}
where the $\lambda_{i_{1},i_{2},...,i_{N}}$ are the eigenvalues, in decreasing order, of the Hermitian matrix as:}
\begin{eqnarray}\label{R}
{\bf R=\sqrt{\sqrt{\rho}\widetilde{\rho}\sqrt{\rho}}.}
\end{eqnarray}
Note that each $\lambda_{i_{1},i_{2},...,i_{N}}$ is a non-negative real number and that the all of $i_{1},...,i_{N}$ cannot be zero simultaneously. The proof of this formula is similar to the proof of the Wootters in \cite{Wootters}. The general algorithm is as follows:\\\\
First, find a complete set of orthogonal eigenvectors $|v_{i_{1},i_{2},...,i_{N}}\rangle$ corresponding to the nonzero eigenvalues of $\rho$ and subnormalize these vectors so that $\langle v_{i_{1},i_{2},...,i_{N}}|v_{i_{1},i_{2},...,i_{N}}\rangle$ is equal to the i-th eigenvalue: $|v_{i_{1},i_{2},...,i_{N}}\rangle=\sqrt{p_{i_{1},i_{2},...,i_{N}}}|\psi_{i_{1},i_{2},...,i_{N}}\rangle$. So the density matrix of $\rho$ is written as:
 $$\rho=\sum_{i_{1},i_{2},...,i_{N}=0}^{1}|v_{i_{1},i_{2},...,i_{N}}\rangle\langle v_{i_{1},i_{2},...,i_{N}}|.$$
 Then we form the symmetric but not necessarily Hermitian matrix of $\tau_{i_{1},i_{2},...,i_{N},j_{1},j_{2},...,j_{N}}=\langle v_{i_{1},i_{2},...,i_{N}}|\widetilde{v}_{j_{1},j_{2},...,j_{N}}\rangle$. The unitary matrix U that diagonalizes the $\tau$, obtain the other decompositions of $\rho$ called it $|x_{i_{1},i_{2},...,i_{N}}\rangle$ such that:
 \begin{eqnarray}\label{x}
   \langle x_{i_{1},...,i_{N}}|\widetilde{x}_{j_{1},...,j_{N}}\rangle=(U\tau U^{T})_{i_{1},i_{2},...,i_{N}j_{1},j_{2},...,j_{N}},
\end{eqnarray}
and:
\begin{eqnarray}\label{second}
 \langle x_{i_{1},...,i_{N}}|\widetilde{x}_{j_{1},...,j_{N}}\rangle=\lambda_{i_{1},i_{2},...,i_{N}}\delta_{i_{1}j_{1}}\delta_{i_{2}j_{2}}...\delta_{i_{N}j_{N}}.
\end{eqnarray}
Note that the decompositions $|x_{i_{1},i_{2},...,i_{N}}\rangle$ is related to $|v_{i_{1},i_{2},...,i_{N}}\rangle$ by unitary matrix U. The diagonal
elements of $U\tau U^{T}$ can always be made real and nonnegative, in which case they are the square roots of the
eigenvalues of $\tau\tau^{*}$ that are the same as the eigenvalues of $R$. On the other hand, if we consider the other decompositions of $\rho$ as:
\begin{widetext}
\begin{eqnarray}
|y_{0,0,...,0}\rangle=|x_{0,0,...,0}\rangle\quad\quad\quad\quad\quad\quad\quad\quad\quad\cr\cr
|y_{j_{1},j_{2},...,j_{N}}\rangle=i|x_{j_{1},j_{2},...,j_{N}}\rangle, for\,\,\,\, j_{1},j_{2},...,j_{N}\neq0 \,\, simultaneously,
\end{eqnarray}
\end{widetext}
then it can be proved that the average polynomial invariant of degree 2 for even-N qubits mixed states is as follows \cite{Wootters}:
\begin{widetext}
\begin{eqnarray}\label{b}
   \sum_{i_{1},i_{2},...,i_{N}}\langle y_{i_{1},i_{2},...,i_{N}}|y_{i_{1},i_{2},...,i_{N}}\rangle\frac{\langle y_{i_{1},i_{2},...,i_{N}}|\widetilde{y}_{i_{1},i_{2},...,i_{N}}\rangle}{\langle y_{i_{1},i_{2},...,i_{N}}|y_{i_{1},i_{2},...,i_{N}}\rangle}\cr
   =\sum_{i_{1},i_{2},...,i_{N}}\langle y_{i_{1},i_{2},...,i_{N}}|\widetilde{y}_{i_{1},i_{2},...,i_{N}}\rangle\cr
   =\lambda_{0,0,...,0}-\sum_{i_{1},i_{2},...,i_{N}}\lambda_{i_{1},i_{2},...,i_{N}}=C(\rho),
\end{eqnarray}
\end{widetext}
where with no loss of generality, we consider that $\lambda_{0,0,...,0}$ is greater than the others. So the decompositions of $\{y_{i_{1},i_{2},...,i_{N}}\}$ satisfy the polynomial invariant of degree 2 for any even-N qubits mixed state. Now, as the same way of the \cite{Wootters}, can be shown that for any other decompositions $\{|z_{i_{1},i_{2},...,i_{N}}\rangle\}$ of $\rho$ which related to $\{|y_{i_{1},i_{2},...,i_{N}}\rangle\}$ by unitary matrix $V$ as:
$$|z_{i_{1},i_{2},...,i_{N}}\rangle=\sum_{j_{1},j_{2},...,j_{N}}V^{*}_{i_{1},i_{2},...,i_{N}j_{1},j_{2},...,j_{N}}|y_{j_{1},j_{2},...,j_{N}}\rangle,$$
the average polynomial invariant of degree 2 for any even-N qubits mixed state is greater than $C(\rho)$:
\begin{widetext}
\begin{eqnarray}\label{de}
   \sum_{i_{1},i_{2},...,i_{N}}|\langle z_{i_{1},i_{2},...,i_{N}}|\widetilde{z}_{i_{1},i_{2},...,i_{N}}\rangle|
   =\sum_{i_{1},i_{2},...,i_{N}}|(VYV^{T})_{ii}|\cr
   \geq\lambda_{0,0,...,0}-\sum_{i_{1},i_{2},...,i_{N}}\lambda_{i_{1},i_{2},...,i_{N}}=C(\rho).
\end{eqnarray}
\end{widetext}
So there is no decomposition of $\rho$ that can be lower than $C(\rho)$.\\
As yet, we consider that $\lambda_{0,0,...,0}\geq\sum_{i_{1},i_{2},...,i_{N}}\lambda_{i_{1},i_{2},...,i_{N}}$.\\
Now, if consider the case $\lambda_{0,0,...,0}-\sum_{i_{1},i_{2},...,i_{N}}\lambda_{i_{1},i_{2},...,i_{N}}\leq0$, can be shown that the polynomial invariant of degree 2 for any even-N qubits mixed state for this case, is zero. In this case, there are the phase factors $\theta_{j_{1},...,j_{N}}$ that can be chosen such that the polynomial invariant of degree 2 for any even-N qubits mixed state is zero for the following set:
\begin{widetext}
\begin{eqnarray}\label{zero}
   |z_{i_{1},i_{2},...,i_{N}}\rangle=\frac{1}{\sqrt{2^{N}}}\sum_{j_{1},j_{2},...,j_{N}}(-1)^{i_{1}j_{1}+...+i_{N}j_{N}}e^{i\theta_{j_{1},...,j_{N}}}|x_{j_{1},...,j_{N}}\rangle.
 \end{eqnarray}
 \end{widetext}
Note that such of these phase factors satisfy: $\sum_{j_{1},...,j_{N}=0}^{1}e^{2i\theta_{j_{1},...,j_{N}}}\lambda_{j_{1},...,j_{N}}=0$.\\ The convex combinations of the set of $|z_{i_{1},i_{2},...,i_{N}}\rangle\langle z_{i_{1},i_{2},...,i_{N}}|$ form zero-polytope that is the convex set of density matrices by vanishing the degree 2 polynomial invariant. Outside
there is no other state with zero degree 2 polynomial invariant.\\


\section{Example}
In this section we compare our formula of polynomial invariant of degree 2 for the even-N qubits mixed states and genuine multipartite concurrence \cite{Rafsanjani} by an example. A system consisting of N qubits is said to has genuine multipartite entanglement if each qubit is entangled
with all of the other qubits and not only to some of them. A pure N-qubit state $|\psi\rangle$, is called biseparable if it is separable under some bipartition. In other words if the pure state $|\psi\rangle$ can be written as $|\psi\rangle=|\phi_{1}\rangle\otimes|\phi_{2}\rangle$, the state is biseparable. We denote this biseparable pure state as $|\psi^{B.S}\rangle$. A mixed state is biseparable if it can be written as $\rho^{B.S}=\sum_{k}p_{k}|\psi_{k}^{B.S}\rangle\langle\psi_{k}^{B.S}|$, that $|\psi_{k}^{B.S}\rangle$ might be biseparable. One of the measures to compute the genuine multipartite entanglement is genuine multipartite concurrence that for a pure state $|\psi\rangle$ is defined as:\\
\begin{eqnarray}\label{GMP}
C_{GM}(|\psi\rangle)=min_{\lambda\in\tau}\sqrt{2}\sqrt{1-Tr(\rho_{A_{\lambda}}^{2})},
\end{eqnarray}
where  $\tau$ represents the set of all possible bi-partitions $\{A_{\lambda}|B_{\lambda}\}$, and $\rho_{A_{\lambda}}$ is the reduced density matrix:$\rho_{A_{\lambda}}=Tr_{B_{\lambda}}(|\psi\rangle\langle\psi|)$. For the mixed state $\rho=\sum_{i}p_{i}|\psi^{i}\rangle\langle\psi^{i}|$, this entanglement measure is defined as the average pure-state GM concurrence:
\begin{eqnarray}\label{GMP}
C_{GM}(\rho)=inf_{\{p_{i},|\psi_{i}\rangle\}}\sum_{i}p_{i}C_{GM}(|\psi^{i}\rangle),
\end{eqnarray}
where minimization is over all possible pure state decompositions of $\rho$.\\
  In \cite{Rafsanjani}, Rafsanjani and co-workers finded an algebraic formula for the N-partite concurrence of N qubits in an X matrix. The general form of X density matrix is given by:
  \begin{eqnarray}\label{example}
\rho  = \sum\limits_{{i_1},...,{i_N} = 0}^1 {(p_{{i_1}...{i_N}}^ + } \left| {\psi _{{i_1}...{i_N}}^ + } \right\rangle \left\langle {\psi _{{i_1}...{i_N}}^ + } \right| +\cr
 p_{{i_1}...{i_N}}^ - \left| {\psi _{{i_1}...{i_N}}^ - } \right\rangle \left\langle {\psi _{{i_1}...{i_N}}^ - } \right|)
\end{eqnarray}
where:
\begin{widetext}
\[\begin{array}{l}
\left| {\psi _{{i_1}...{i_N}}^ + } \right\rangle  = \cos (\frac{{{\theta _{{i_1}...{i_N}}}}}{2})\left| {{i_1}...{i_N}} \right\rangle  + {e^{i{\varphi _{{i_1}...{i_N}}}}}\sin (\frac{{{\theta _{{i_1}...{i_N}}}}}{2})\left| {{{\overline i }_1}...{{\overline i }_N}} \right\rangle \\
\left| {\psi _{{i_1}...{i_N}}^ - } \right\rangle  =  - \sin (\frac{{{\theta _{{i_1}...{i_N}}}}}{2})\left| {{i_1}...{i_N}} \right\rangle  + {e^{i{\varphi _{{i_1}...{i_N}}}}}\cos (\frac{{{\theta _{{i_1}...{i_N}}}}}{2})\left| {{{\overline i }_1}...{{\overline i }_N}} \right\rangle
\end{array}\]
\end{widetext}
that $\overline{i}=i+1$ in modulo 2 arithmetic. Here, we consider any even-N qubits of general form of X density matrix and compute $C(\rho)$   by means of Eq.(\ref{HHHHHH}). The result is as following form:
\begin{eqnarray}\label{result1}
C(\rho ) = \max \{ 0,\left| {{z_{{i_1}...{i_N}}}} \right| -\sum_{j\neq i} \sqrt {{a_{{j_1}...{j_N}}}{b_{{j_1}...{j_N}}}} \},
\end{eqnarray}
where $a_{{j_1}...{j_N}}$ and $b_{{j_1}...{j_N}}$  are  $2^{N-1}$ diagonal elements of density matrix, and $z_{{i_1}...{i_N}}$  are $2^{N-1}$  off-diagonal elements of this density matrix:
\[\begin{array}{l}
{a_{{j_1}...{j_n}}} = p_{_{{j_1}...{j_N}}}^ + {\cos ^2}(\frac{{{\theta _{{j_1}...{j_N}}}}}{2}) + p_{_{{j_1}...{j_N}}}^ - {\sin ^2}(\frac{{{\theta _{{j_1}...{j_N}}}}}{2})\\
{b_{{j_1}...{j_n}}} = p_{_{{j_1}...{j_N}}}^ - {\cos ^2}(\frac{{{\theta _{{j_1}...{j_N}}}}}{2}) + p_{_{{j_1}...{j_N}}}^ + {\sin ^2}(\frac{{{\theta _{{j_1}...{j_N}}}}}{2})\\
{z_{{i_1}...{i_n}}} = \frac{1}{2}{e^{ - i{\varphi _{{i_1}...{i_N}}}}}\sin ({\theta _{{i_1}...{i_N}}})(p_{_{{i_1}...{i_N}}}^ +  - p_{_{{i_1}...{i_N}}}^ - )
\end{array}\]
Eq.(\ref{result1}) is in full agreement with the $C_{GM}$ in \cite{Rafsanjani} and show that our formula can compute the genuine multipartite entanglement of this type of density matrices.\\

In summary, one of the challenges in present quantum information
theory is the quantification of multipartite entanglement
in mixed states that can be obtained via the convex-roof
extension. The convex roof is obtained by minimizing the average
entanglement of a given mixed state over all possible decompositions
of that state into pure states. We have focused here on the polynomial invariant of degree 2 for any even-N qubits state which is zero for any odd-qubits state. Then we have found the explicit formula for computing this invariant for any even-N qubits mixed quantum state. Finally we have tested this formula on general form of even-N qubits mixed state that is in $X$ matrix form and have found complete
agreement with GM concurrence. So our formula makes possible the easy evaluation of the genuine multipartite entanglement for X density matrices.\\\\


\end{document}